\documentclass[%
 reprint,
 amsmath,amssymb,
aps
]{revtex4-2}

\usepackage{graphicx}
\usepackage{dcolumn}
\usepackage{bm}
\usepackage{physics}
\usepackage{xcolor}

\usepackage{xr}
\makeatletter
\newcommand*{\addFileDependency}[1]{
  \typeout{(#1)}
  \@addtofilelist{#1}
  \IfFileExists{#1}{}{\typeout{No file #1.}}
}
\makeatother

\newcommand*{\myexternaldocument}[1]{%
    \externaldocument{#1}%
    \addFileDependency{#1.tex}%
    \addFileDependency{#1.aux}%
}

\myexternaldocument{appendix}

\renewcommand{\[}{\begin{equation}\begin{aligned}} 
\renewcommand{\]}{\end{aligned}\end{equation}}

\newcommand*{\kbt}{k_{\textup{B}}T}

\begin{document}


\title{Immune cells use active tugging forces to distinguish affinity and accelerate evolution}

\author{Hongda Jiang and Shenshen Wang}
 \email{shenshen@physics.ucla.edu}
 \affiliation{Department of Physics and Astronomy, University of California, Los Angeles, Los Angeles, CA 90095}


%
%

\date{\today}

\begin{abstract}
Cells are known to exert forces to sense their physical surroundings for guidance of motion and fate decisions. Here, we propose that cells might do mechanical work to drive their own evolution, taking inspiration from the adaptive immune system.
Growing evidence indicates that immune B cells -- capable of rapid Darwinian evolution -- use cytoskeletal forces to actively extract antigen from other cells' surface.
To elucidate the evolutionary significance of force usage, we develop a theory of tug-of-war antigen extraction that maps receptor binding characteristics to clonal reproductive fitness, revealing physical determinants of selection strength. 
This framework unifies mechanosensing and affinity-discrimination capabilities of evolving cells: pulling against stiff antigen tethers enhances discrimination stringency at the expense of absolute extraction. 
As a consequence, active force usage can accelerate adaptation but may also cause extinction of cell populations, resulting in an optimal range of pulling strength that matches molecular rupture forces observed in cells.
Our work suggests that nonequilibrium, physical extraction of environmental signals can make biological systems more evolvable at a moderate energy cost. 
\end{abstract}

\pacs{Valid PACS appear here}
\maketitle


\section{Introduction}
The ability of cells to sense and respond to mechanical forces is critical to many important processes in biology, from embryonic development~\cite{puliafito2012} and wound healing~\cite{ng2012} to migration of cancer cells~\cite{labernadie2017} and immune recognition~\cite{depoil2014}. Over the past decades, much research has focused on understanding how physical forces, applied to cell-cell or cell-material interfaces, coordinate the movement of cells as a collective~\cite{trepat2018} or guide the fate decisions of individual cells~\cite{engler2006}. However, rarely is a link drawn between active force exertion on cell-cell contact and evolutionary outcome of a cell population. Here we propose such a link in the context of antigen recognition by immune cells, demonstrating how physical dynamics and rapid evolution interplay across scales to shape the emergent responses. 

An adaptive immune response begins with the activation of B and T cells, mediated by specific binding of their unique surface receptors to antigens presented on other cell surface. The response ends with the formation of immune memory, composed of diverse clones expressing receptors with varying affinity for the encountered pathogen and confer protection against future reinfections. In between, an accelerated Darwinian evolutionary process takes place in dynamic open microenvironments known as germinal centers (GCs)~\cite{victora2012}, where B cells compete and evolve to produce high affinity antibodies, i.e. membrane-detached B cell receptors (BCRs). Within a few weeks of an infection, antibody affinities can increase by a thousand fold~\cite{eisen1964}. However, this evolutionary process of affinity maturation (AM) eventually saturates. Notably, maximum affinities evolved \textit{in vivo} are orders of magnitude lower than those realized by directed evolution \textit{in vitro}~\cite{boder2000, poulsen2011}. The exact origin of affinity ceiling~\cite{batista1998} and its functional implications remain unknown.

New technological advances in live-cell imaging and sensitive force probes have revealed a strongly physical and inherently nonequilibrium picture of immune recognition: via cell-cell contact, a B cell not only engages but also extracts antigen from the antigen presenting cell (APC) by applying mechanical pulling forces generated by its contractile cytoskeleton~\cite{ natkanski2013, nowosad2016, kwak2018}. Experiments and computational studies have demonstrated a role of cytoskeletal flows in patterning the contact domain in T cell synapse~\cite{qi2001, hartman2009, weikl2004}. Only recently, theoretical work~\cite{knevzevic2018} has found pulling forces to be critical to creating and maintaining the multifocal synaptic pattern observed in GC/evolving B cells~\cite{nowosad2016}, in stark contrast to the bull’s eye pattern seen in naive (antigen-inexperienced) and memory (differentiated) cells. However, the question remains as how, and why, evolving B cells use active tugging forces to physically extract antigen through cellular interfaces. Despite conceptual proposals~\cite{batista1998, desikan2021} that affinity maturation may be limited by factors of a physical origin, no quantitative framework is yet available for verifying or falsifying this idea. 

Two broad classes of mechanisms have been proposed to explain the remarkable sensitivity and specificity of immune recognition:  
Kinetic proofreading~\cite{hopfield1974, ninio1975} reduces error rates of discrimination through serial amplification of small differences, via a cascade of biochemical reactions that creates a controlled delay. Mechanical proofreading~\cite{brockman2019}, on the other hand, employs pico-newton (pN) molecular forces to enhance fidelity of information transfer by eliciting catch-bond behavior~\cite{marshall2003,liu2014}, mediating receptor clustering~\cite{van2011,manz2011t}, or inducing conformational changes~\cite{savir2007,alon2007}. However, existing models cannot address potential functional advantage of physical acquisition of antigen via nonequilibrium bond rupture. Furthermore, \textit{in vivo} experiment~\cite{gitlin2014} indicates that reproductive fitness of a B cell is primarily determined by the total amount of antigen it acquires from the APC, suggesting that force-dependent efficiency of antigen extraction provide a map from intrinsic binding quality to a selectable phenotype. 

Here, we develop a theory of stochastic antigen extraction and elucidate the role of active tugging forces in affinity discrimination, in light of adaptive evolution of immune responses. By describing how mechanical stress propagates through a chain of binding interfaces to deform the combined free energy landscape, this physical theory recapitulates and unifies mechanosensing~\cite{spillane2017} and affinity-discrimination~\cite{natkanski2013} capabilities of immune cells: pulling against stiff APCs reduces the absolute level of antigen extraction but enhances the stringency of discrimination between similar BCR affinities. Further, this model predicts that force-induced landscape deformation stretches the response curve and widens the discrimination range; such range expansion is even more significant if force magnitude ramps up over time. 

To allow microscopic interpretation of the mapping, we describe how to extract intrinsic parameters of the multi-dimensional free energy profile -- from data collapse of rupture force histograms obtainable from dynamic force spectroscopy. Finally, by subjecting the binding phenotype to \textit{in silico} evolution, we find that stronger pulling raises the affinity ceiling and accelerates adaptation, but at a risk of population extinction. Remarkably, the preferred force magnitude ($10$--$20\mathrm{pN}$) --- predicted to balance population survival and adaptation --- appears to match the range of rupture forces measured by single-molecule pulling experiment~\cite{natkanski2013} and DNA-based tension sensors in live B cells~\cite{spillane2017}. In all, this work makes a first step toward a quantitative framework of cell-mediated evolution of molecular recognition, revealing the impact of active forces and physical dynamics of cells on selection pressure. We expect the approach and principles to have broad relevance to biological recognition systems, where the efficiency of signal acquisition by physical means dictates the selective advantage of competing cells.

\begin{figure*}[t]\label{fig1}
    \includegraphics[width=0.86\textwidth]{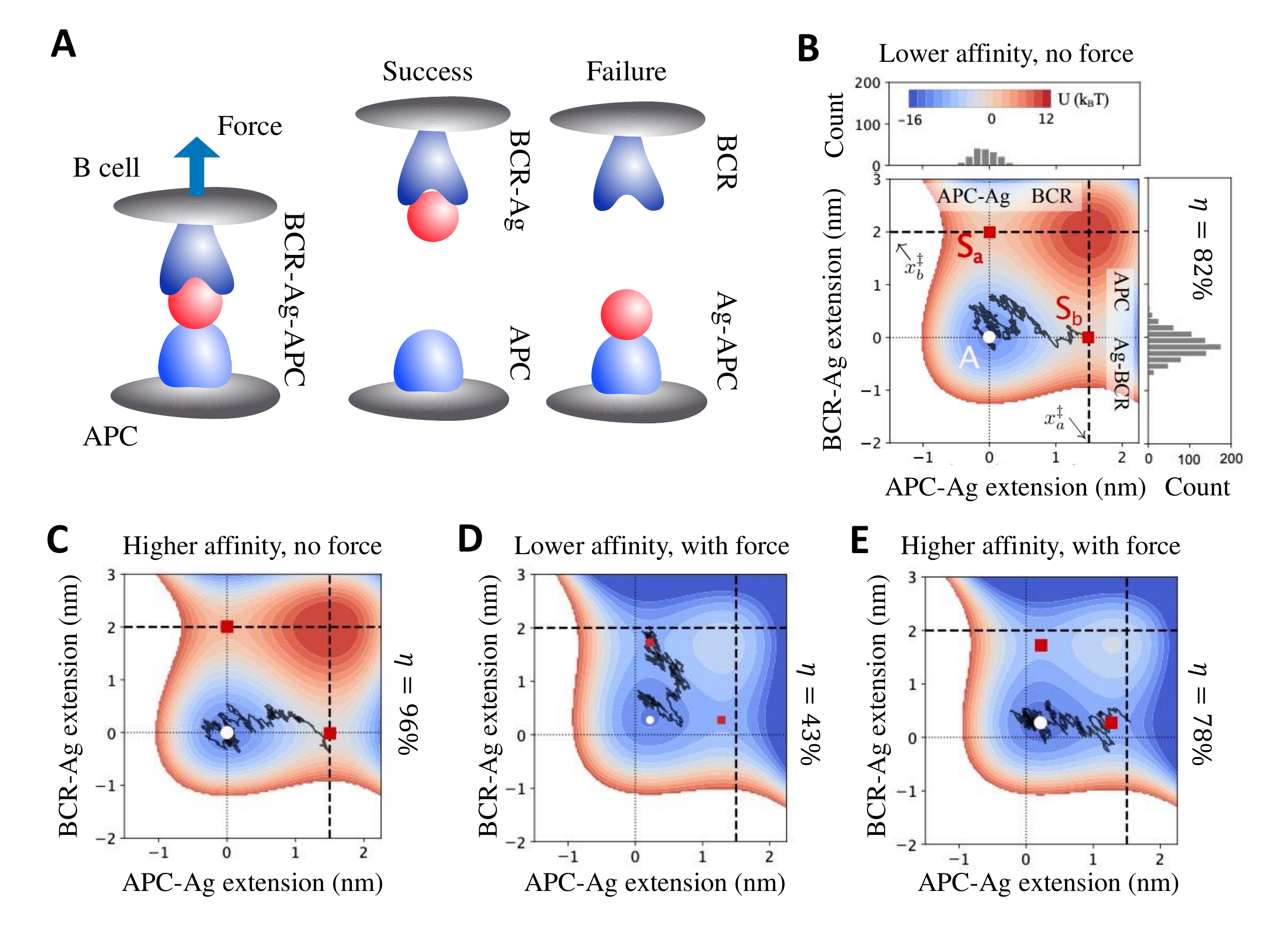}
    \caption{B cells acquire antigen and discriminate receptor affinity through a molecular tug of war. (A) Schematic illustration of an extraction attempt in a tug-of-war configuration: pulling force exerted by a B cell stretches a BCR-Ag-APC complex on the surface of an antigen presenting cell (APC), leading to one of two outcomes --- a sooner rupture of the Ag-APC bond makes a success, whereas a faster dissociation of the BCR-Ag bond yields a failure. The Ag-APC bond coarse-grains potentially complex interactions including Ag-tether association and strength of the APC membrane; a short-lived Ag-APC bond under force can be due to a weak Ag-tether bond or a soft APC membrane, which we do not distinguish. (B) System state can be specified in terms of the extension of the Ag-APC bond and that of the BCR-Ag bond. Stochastic Ag extraction occurs via thermal escape from the bound state (attractor at A) over one of the activation barriers (saddles at $S_a$ and $S_b$) in the binding free energy landscape (color coded). A complex ruptures as soon as the trajectory (black trace) hits one of the absorbing boundaries located at rupture lengths $x_a=x_a^\ddagger$ and $x_b=x_b^\ddagger$ (dashed lines). In the absence of force, a relatively high BCR affinity ($\Delta G_b^\ddagger=12\kbt$) leads to a high chance of Ag extraction ($\eta=82\%$); the grey histograms show the distribution of exit position at each boundary, i.e. the extension of the remaining bond when the other breaks. (C) An increase in BCR affinity ($12\kbt\rightarrow 14\kbt$) leads to a moderate fractional increase in the already high extraction likelihood ($82\%\rightarrow 96\%$). (D, E) Tugging force ($F=20$pN) deforms the binding free energy landscape, displacing the attractor and two saddles, as well as lowering two barriers by different amounts. Such deformation reduces the absolute level of extraction ($82\%\rightarrow 43\%$, $96\%\rightarrow 78\%$) but greatly enhances the contrast between similar affinities ($82\%$ vs $96\% \rightarrow 43\%$ vs $78\%$). Parameters: $\Delta G_a^\ddagger=12\kbt, x_a^\ddagger=1.5{\rm nm}, x_b^\ddagger=2{\rm nm}$.       
    } 
\end{figure*}

\section{Theory of antigen extraction}

Antigen (Ag) extraction occurs via a molecular tug of war under active pulling forces (illustrated in Fig.~\ref{fig1}A): productive binding of BCRs and antigens tethered to the surface of APCs triggers B cell intracellular signaling and generation of contractile forces, which pull on chains of protein complexes that connect a B cell to the APC~\cite{batista2009, spillane2018}. Along the chain, tugging force propagates through a series of binding interfaces, altering the extension of coupled molecular bonds in the pulling direction. Acquisition of an antigen requires rupturing its membrane tether. Thus, the tug-of-war setting of antigen extraction implements a comparison of binding quality via competitive rupture of tugging and tethering complexes.

As a first step, we consider a coarse-grained description of BCR-Ag-APC three-body complexes, in which the Ag-APC attachment may involve multivalent binding in an intricate geometry (e.g. an antibody-coated antigen cluster tethered to the APC membrane by multiple complement receptors) whose overall lifetime sets the tether strength. For simplicity, we assume independent complexes subject to equal pulling stress, a mean-field scenario consistent with the observation that traction force applied to a BCR cluster scales with its size~\cite{tolar2014, wang2018}. In a three-body complex, antigen movement couples bond extensions. Hence, system dynamics proceeds on a combined free energy surface, deformed by pulling force, over a 2D state space spanned by $x_a$ and $x_b$, the extension of the Ag-APC (tethering) bond and the BCR-Ag (tugging) bond, respectively (see example profiles in Fig.~\ref{fig1}B-E):  
\begin{equation}\label{potential}
U(x_a, x_b; t)=U_a(x_a)+U_b(x_b)+V_{\rm pull}(x_a+x_b; t).\nonumber
\end{equation}
Here $V_{\rm pull}(x; t)=-F(t)x$ describes landscape deformation caused by pulling force $F(t)$. The intrinsic free energy profile $U_a(x_a)$ ($U_b(x_b)$) has a potential well at zero extension and a barrier of height $\Delta G_a^\ddagger$ at the rupture length $x_a=x_a^\ddagger$ (of height $\Delta G_b^\ddagger$ at $x_b=x_b^\ddagger$). The combined surface $U$ has an attractor $A$ and two saddle points $S_a$ and $S_b$ (Fig.~\ref{fig1}B). Applied forces lower both potential barriers (but to different extents) and displace the attractor and saddles, resulting in shorter minimum-to-barrier distances and smaller curvatures at the well and barriers (Fig.~\ref{fig1}D-E).

Antigen extraction is inherently stochastic, because rupture of a chain of adhesive bonds occurs through thermally aided escape from the bound state over one of the activation barriers~\cite{hanggi1990}. Consider a 2D description of the rupture process along coupled reaction coordinates $x_a$ and $x_b$. Extraction dynamics is governed by coupled Langevin equations describing the motion of antigen and BCR, respectively (see SI for details): 
\begin{eqnarray}\label{langevin}
\gamma_a \dot x_a &=& -U_a'(x_a) + U_b'(x_b)+\xi_a,\\\nonumber
\gamma_b (\dot x_a+\dot x_b) &=& -U_b'(x_b)+F+\xi_b.
\end{eqnarray}
Here $\xi_a$ and $\xi_b$ are random forces caused by collision with particles in the fluid, yielding a distribution of rupture forces; $\langle\xi_i\rangle=0$, $\langle\xi_i(t)\xi_j(t')\rangle=2k_BT\gamma_i\delta_{ij}\delta(t-t')$, with $i,j=a,b$. Frictional coefficients $\gamma_a$ and $\gamma_b$ set the relaxation timescale. 
These equations of motion reflect a balance of frictional forces, elastic forces ($-U_a'(x_a)$ and $-U_b'(x_b)$), pulling force $F$, and random forces. 

Intuitively, a molecular tug of war acts to compare the strength, measured by lifetime or rupture force, of the BCR-Ag bond and the Ag-APC bond under pulling stress. Therefore, estimating the probability of antigen extraction boils down to comparing the first passage time to reaching either absorbing boundary, i.e., exceeding one of the bond rupture lengths ($x_a^\ddagger$ and $x_b^\ddagger$). The probability of successful extraction, $\eta$, can thus be expressed as the chance by which the lifetime of the BCR-Ag bond, $t_b$, exceeds that of the Ag-APC bond, $t_a$, given their respective lifetime distributions $p_b(t)$ and $p_a(t)$; in the limit of high activation barriers, a simple form results

\begin{figure*}[t]
\begin{center}
\includegraphics[width=0.85\textwidth]{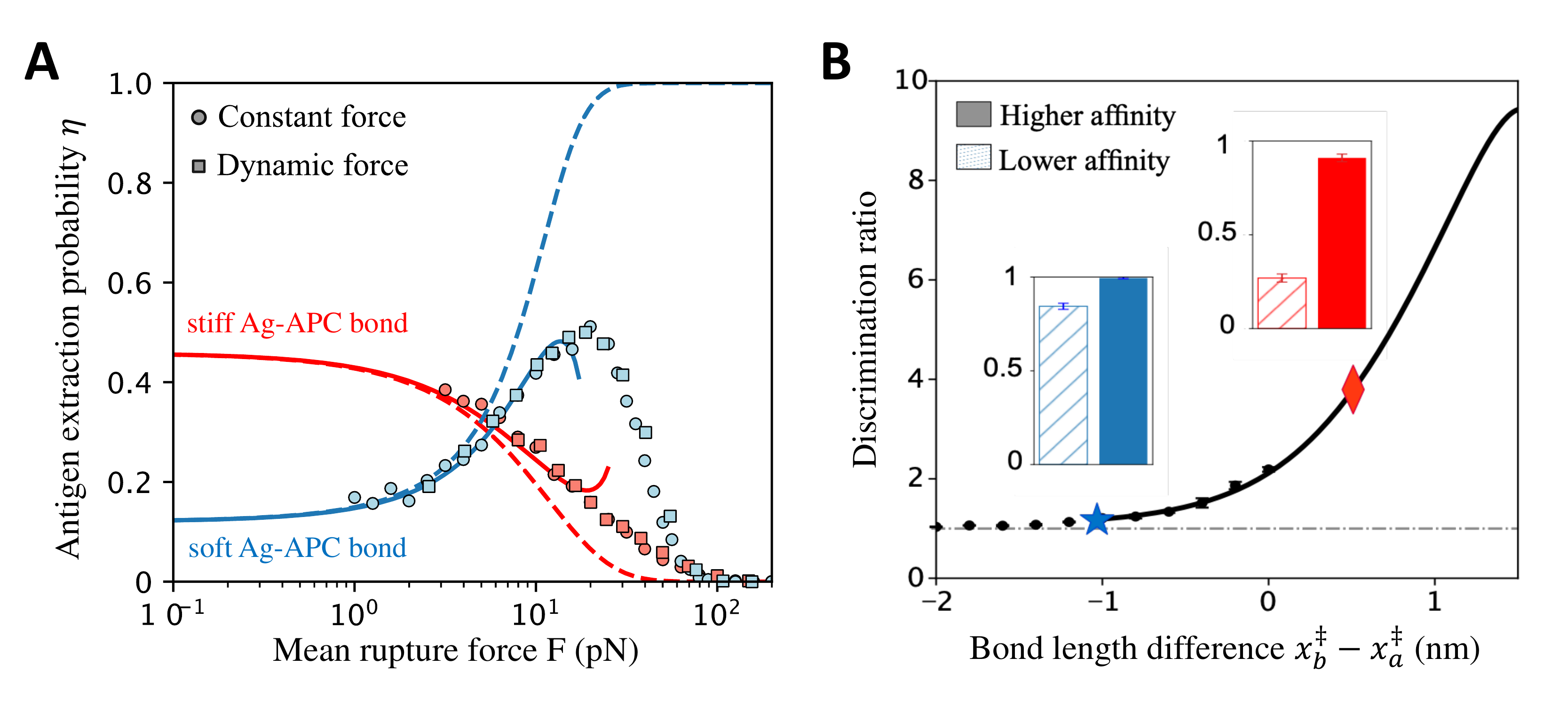}
\caption{
Tug-of-war antigen extraction enables mechanosensing and affinity discrimination. 
The slip-bond characteristic of BCR-Ag interaction underlies reduced extraction and enhanced discrimination stringency when B cells pull against stiff antigen tethers. 
(A) Antigen extraction can increase or decrease due to pulling, depending on whether the APC-Ag bond is stiffer (red, $x_a^\ddagger < x_b^\ddagger$) or softer (blue, $x_a^\ddagger > x_b^\ddagger$) than the BCR-Ag bond. Mean rupture force is predictive of antigen extraction, whether the pulling force is constant (circle) or ramping up over time (square; loading rate: $1-10^6$ pN/s). Brownian dynamics simulations (symbols, 1000 runs each) show excellent match with the constant-force predictions based on Eqs.~\ref{eta_high_barrier} and \ref{lifetime} (solid lines). Bell's model (dashed lines) works for low force but fails already under modest force below 10 pN. 
$\Delta G_a^\ddagger=\Delta G_b^\ddagger=10\kbt$; red: $x_a^\ddagger=1.5{\rm nm}$, blue: $x_a^\ddagger=3{\rm nm}$.
(B) Discrimination stringency, measured by the ratio of extraction probability between B cells with different affinities ($15\kbt$ vs $19\kbt$), increases with the difference in bond lengths $x_b^\ddagger-x_a^\ddagger$; a larger difference indicates a stiffer Ag-APC bond. The solid line is theory and black dots are averages over repeated simulations. Insets show statistics of extraction levels at two stiffness values indicated with a red diamond and a blue star. $\Delta G_a^\ddagger=14\kbt$, $F=20{\rm pN}$. 
}
\label{fig2}
\end{center}
\end{figure*}

\begin{equation}\label{eta_define}
\eta = P(t_b>t_a)=\int_0^{\infty}dt p_a(t)\int_{t}^{\infty}dt'p_b(t').
\end{equation}
Here the second integral represents the survival probability of the BCR-Ag bond until at least time $t$ when the Ag-APC bond breaks. 
Note that $p_a(t)$ and $p_b(t)$ are first passage time distributions calculated with the competing bond being longer lived (i.e. treated as a reflective boundary).
In general, $\eta$ should be found from an integrated probability flux in the 2D state space, conditioned on exiting through the relevant absorbing boundary (here, rupture of the Ag-APC bond). Yet, high activation barriers yield a separation of timescales between relaxation around the attractor and arrival at the transition state, allowing $p_a(t)$ and $p_b(t)$ to factorize.

Through affinity maturation, the activation barrier $\Delta G_b^\ddagger$ evolves considerably over time, whereas the bond length $x_b^\ddagger$ changes to a lesser extent. As shown using single-molecule force spectroscopy, the minimum-to-barrier distance ($x_b^\ddagger\sim0.9$nm) remains similar among mutated variants of recombinant antibody fragments~\cite{morfill2007}. Hereafter affinity discrimination refers to distinguishing the height of activation barrier. 

Below we elucidate the impact of force on affinity discrimination, identify determinants of its stringency and operation range, and demonstrate ways in which energy-consuming physical extraction of antigen may confer evolutionary benefits to immune adaptation.

\section{Results}

\subsection{Unifying mechanosensing and affinity discrimination}
T cells employ multiple modes of mechanical proofreading to stabilize receptor binding and achieve exquisite specificity of self-nonself discrimination. These include catch-bond behavior during activation~\cite{liu2014} and negative selection~\cite{hong2018}, conformational changes of adhesion molecules~\cite{comrie2015} and receptor clustering~\cite{cai2017}. In contrast, B cells apply forces to segregate and rupture BCR-Ag clusters and individual antibody-antigen interactions exhibit a slip-bond character~\cite{schwesinger2000}, i.e., lifetime reduction under force, hinting at a different mechanism of affinity discrimination and a distinct functional goal via antigen extraction.   

How do tugging forces modulate antigen extraction from the APC?  
Under modest constant pulling forces, distributions of bond lifetime are nearly exponential. The resulting extraction probability is simply 
\begin{equation}
\label{eta_high_barrier}
\eta =\frac{1}{1+\tau_a/\tau_b},
\end{equation}
which only depends on the ratio of mean bond lifetimes, $\tau_a$ for the Ag-APC bond and $\tau_b$ for the BCR-Ag bond. In other words, chance of extraction is determined by the relative rate of escape from the bound state across two saddles (rupture lengths). 
For diffusive dynamics, one can calculate the lifetimes using Langer's multidimensional generalization of Kramers theory~\cite{Langer1969} to obtain 
\begin{equation}
\label{Langer}
\eta=\left[1+\frac{\tau_a^{+}}{\tau_b^{+}}\sqrt{\frac{|\mathrm{det}\mathbf{H}_{S_a}|}{|\mathrm{det}\mathbf{H}_{S_b}|}} e^{\beta(U_{S_a}-U_{S_b})}\right]^{-1}.  
\end{equation}
Here $\tau_i^{+}$ is the characteristic time to escape from a saddle point and $\mathbf{H}$ is the Hessian matrix at the transition state. 
Pulling force affects antigen extraction in two ways: it modulates the gap between apparent activation energies, $U_{S_a}-U_{S_b}$, in the Arrhenius (exponential) factor, and alters the shape of the free energy surface near the saddle points that enters through the prefactor. The extent of such influences depends on the nature of the free energy surface. 
Explicitly, the lifetime can be written in a unified form~\cite{dudko2006} (see SI for derivation)
\begin{equation}\label{lifetime}
\tau_i = \tau_{i0} \exp\left[\beta \Delta G_i^\ddagger \left(1-\frac{vFx^\ddagger_i}{\Delta G_i^\ddagger}\right)^{1/v}\right],
\end{equation}
where $v=2/3$ specifies a linear-cubic potential, $U(x)=(3/2)\Delta G^\ddagger (x/x^\ddagger-1/2) - 2\Delta G^\ddagger (x/x^\ddagger-1/2)^3$, whereas $v=1/2$ corresponds to a cusp-harmonic surface, $U(x)=\Delta G^\ddagger(x/x^\ddagger)^2$ for $x<x^\ddagger$ and $-\infty$ for $x\ge x^\ddagger$. For $v=1$ and for $\Delta G^\ddagger\rightarrow\infty$ independent of $v$, the expression reduces to Bell's phenomenological model~\cite{bell1978}. 
Hence, $U_{S_a}-U_{S_b}=\Delta G_a^\ddagger\left(1-F/f_a\right)^{1/v}-\Delta G_b^\ddagger\left(1-F/f_b\right)^{1/v}$ and $\tau_{a0}/\tau_{b0}\propto\left[(f_b-F)/(f_a-F)\right]^{1/v-1}$, with $f_i=\Delta G_i^\ddagger/vx_i^\ddagger$ being the critical force at which the corresponding barrier vanishes (SI text).
It follows that, if $f_a\approx f_b$, the Arrhenius term (exponential difference in barrier height) dominates the force dependence of $\eta$.

What are the functional consequences of physical extraction of antigen using tugging forces? Force exertion in a tug-of-war configuration renders antigen acquisition sensitive to the physical properties of presenting cells, summarized by the intrinsic affinity $\Delta G_a^\ddagger$ and  rupture length $x_a^\ddagger$ (inverse stiffness) of the antigen tether in our coarse-grained model. 
Specifically, depending on whether the Ag-APC bond is stiffer or softer than the BCR-Ag bond, pulling can either reduce or enhance extraction (Fig.~\ref{fig2}A red vs blue curve).  
This can be understood from the stiffness dependence of barrier reduction induced by modest forces ($F<\min(f_a, f_b)$):
\begin{eqnarray}\label{barrier}
U_{S_b}-U_{S_a}\approx\Delta G_b^\ddagger-\Delta G_a^\ddagger-F(x_b^\ddagger-x_a^\ddagger)\\\nonumber
+\frac{1-v}{2}\left[\frac{(Fx_b^\ddagger)^2}{\Delta G_b^\ddagger}-\frac{(Fx_a^\ddagger)^2}{\Delta G_a^\ddagger}\right].
\end{eqnarray}

To linear order in $Fx^\ddagger$, when the Ag-APC attachment is softer ($x_a^\ddagger>x_b^\ddagger$), force lowers the activation barrier of the BCR-Ag bond rupture by a smaller amount than that of the APC-Ag bond, thus enlarging the barrier gap and promoting extraction (Fig.~\ref{fig2}A blue curve). Conversely, stronger pulling against stiff substrates ($x_a^\ddagger<x_b^\ddagger$) inhibits extraction, due to a reduction in barrier gap (Fig.~\ref{fig2}A red curve). That is, force dependence of extraction flips sign as $x_a^\ddagger$ decreases past $x_b^\ddagger$. Physically, this reflects that in a pulled chain of molecular interactions, force weakens the softer bonds to a larger extent than it does to the stiffer bonds (Fig.~S2). This captures the observed mechanosensing behavior~\cite{spillane2017}: evolving B cells generate strong forces to pull against stiff APCs (e.g. follicular dendritic cells in GCs) and acquire fewer antigens compared to naive cells that apply weak forces. Hence for GC B cells ($x_a^\ddagger < x_b^\ddagger$), tugging force suppresses antigen extraction. Importantly, APCs can modulate their stiffness in response to inflammatory signals~\cite{bufi2015} while B cells can sense and respond to changes in substrate stiffness, suggesting an adaptive mechanism that enables proper responses to complex environmental signals.    

Given the energetic cost of force generation, why do B cells pull to reduce antigen extraction? Interestingly, while pulling against stiff APCs diminishes the absolute level of extraction, the contrast (distinguishability) between similar affinities becomes much enhanced. Fig.~\ref{fig1} B-E shows an example: Without force, BCRs with similar binding affinities yield similarly high extraction.
In contrast, pulling reduces extraction at both affinities but increases the ratio substantially; the chance of extraction is almost doubled with a $2k_\mathrm{B}T$ increase in affinity. 
Essentially, as force reduces the apparent affinity gap between the tugging and tethering bonds, it sensitizes extraction to changes in BCR affinity (to be discussed further below). Mathematically, $\mathrm{d}\log\eta/\mathrm{d}\Delta G_b^{\ddagger}\propto 1-\eta$, indicating higher sensitivity at lower extraction levels. This sensitivity requires stiff APC-Ag associations (Fig.~\ref{fig2}B).

\emph{Regimes of extraction dynamics.} As illustrated in Fig.~\ref{fig2}A, antigen extraction as a function of mean rupture force --- under constant-force (circle) or constant-speed (square) pulling --- exhibits three regimes:

(I) Under sufficiently weak pulling ($F<5$pN), dynamics is primarily governed by noise-assisted exploration of the intrinsic free energy surface where curvatures near the saddle points determine the rupture rate; softer APCs present a smaller curvature at $S_a$ that yields a slower escape from the Ag-APC rupture boundary and hence lower extraction (blue below red curve). In this regime, Bell's model is accurate (dashed and solid lines coincide).  

(II) Intermediate forces ($F\sim 5$--$20$pN) modulate rupture dynamics by deforming the free energy surface, in a way that enables sensing of APC stiffness (Fig.~\ref{fig2}A) and discrimination of BCR affinity (Fig.~\ref{fig2}B). In this functional regime, the Arrhenius factor dominates the force dependence of escape rate and captures the opposite trend as the relative bond length flips sign. However, the value of $\eta$ predicted by Bell's formula deviates from that based on the landscape model, signalling an increasing importance of nonlinear effects associated with landscape deformation, beyond a linear reduction of barrier height (quadratic terms in Eq.~\ref{barrier}). This nonlinearity turns out to be key to expanding the range of distinguishable affinities and sustaining adaptation.  

(III) Once the force is so strong that the barrier to rupture vanishes ($F>20$pN), the attractor and saddles merge and Kramers theory no longer applies; we perform, instead, Brownian dynamics simulations of the coupled Langevin equations (Eq.~\ref{langevin}), shown as symbols in Fig.~\ref{fig2}A. Here, stronger pulling reduces antigen extraction regardless of APC stiffness. This is expected physically: B cell pulling first stretches the BCR-Ag bond, whose relaxation via antigen displacement then deforms the Ag-APC bond, allowing mechanical stress to propagate through the tethering complex after a lag. Under extreme forces, the BCR-Ag bond is so quickly and strongly deformed that it breaks before the Ag-APC bond has time to ``feel" the stress --- antigen hardly moves and extraction vanishes. 
Typical rupture trajectories and extraction statistics under strong pulling confirm the physical intuition (Fig.~S3).  

Therefore, our results suggest that moderate pulling against stiff APCs represents the functional regime of evolving B cells. Most stringent affinity discrimination occurs near but below a critical force where the barrier to rupture vanishes.

\begin{figure}[t]
\begin{center}
\includegraphics[width=0.5\textwidth]{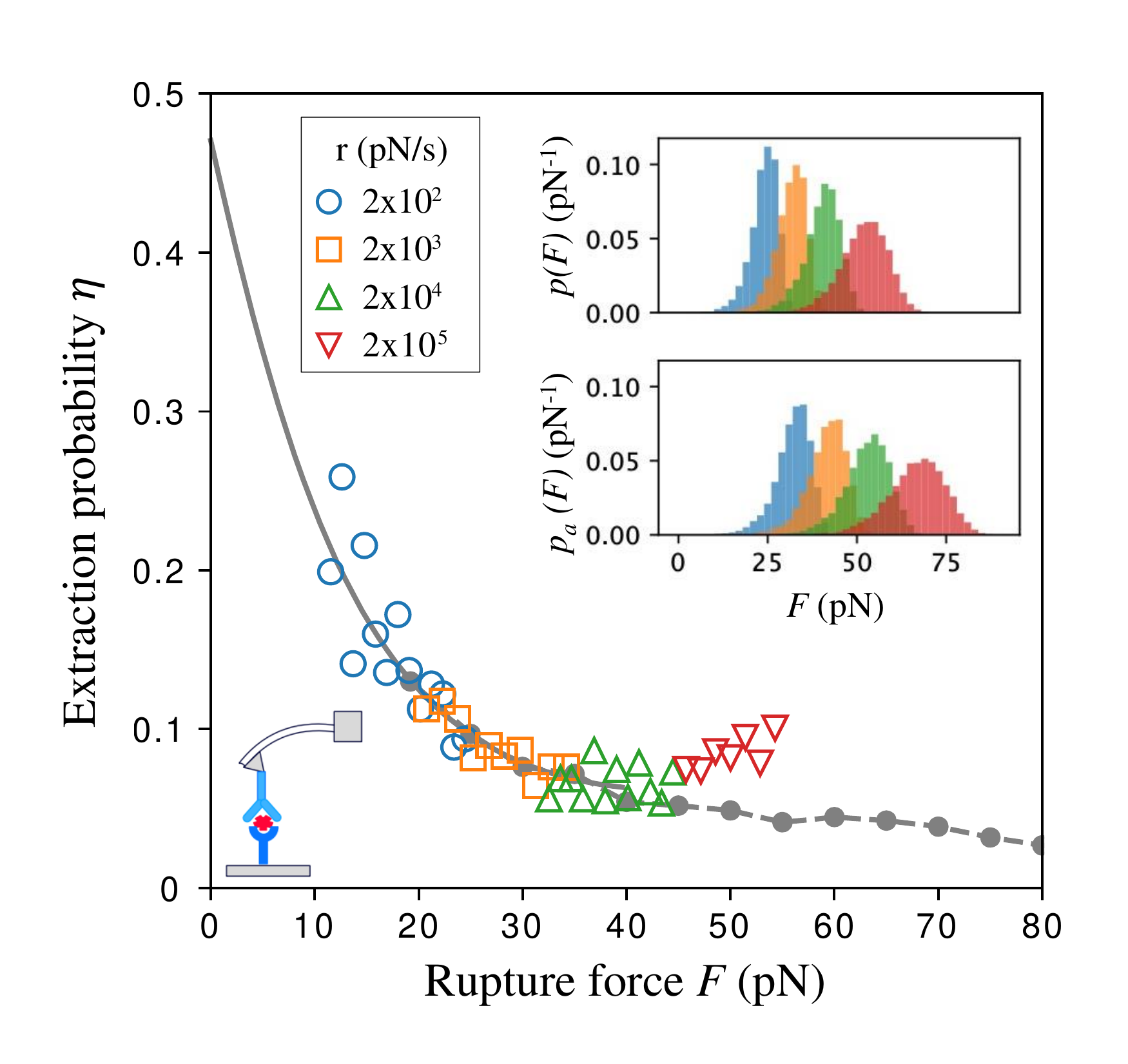}
\caption{
Test of constant-force theory via dynamic-force measurements. 
The theoretical curve of extraction probability under constant force (solid line), based on Eqs.~\ref{eta_high_barrier} and \ref{lifetime}, is able to collapse data (open symbols) obtained by transforming the simulated rupture force histograms (insets) according to Eq.~\ref{reconstruct_eq}. For rupture forces greater than 20pN, we extend the theoretical curve with Brownian dynamics simulations; each filled symbol on the dashed line represents the success rate out of 200 extraction attempts at a given force magnitude. Rupture force histograms cover three decades of loading rate (colors); note that rupture forces of BCR-Ag-APC complexes (upper inset, $p(F)$) tend to have lower values than those of Ag-APC bonds (lower inset, $p_a(F)$), reflecting the fact that $\tau=\tau_a\tau_b/(\tau_a+\tau_b)\leq \min(\tau_a, \tau_b)$, i.e. the shortest-lived bond sets the lifetime and hence rupture force of the entire complex.  
Parameters: $\Delta G_a^{\ddagger}=\Delta G_b^{\ddagger}=20k_{\rm B}T, x_{a}^{\ddagger}=1.5{\rm nm}, x_b^{\ddagger}=2{\rm nm}$. 
}
\label{collapse}
\end{center}
\end{figure}

\subsection{Connecting constant-force theory and dynamic-force measurements}
So far we have treated antigen extraction under constant moderate force, which permits analytical intuition based on the ratio of bond lifetimes.
Notably, several studies have found that B cells tend to generate dynamic forces that may ramp up over time~\cite{kwak2018, kumari2019}. Can we extend the constant-force intuition to understand extraction under dynamic force?

When subject to a steadily ramping force $F(t)=rt$ (e.g. $r=kV$ if a flexible linker of spring constant $k$ pulls at a constant speed $V$), rupture is no longer Poissonian and the transition rate becomes dependent on both the pulling speed and the deformation history of the binding energy landscape. Assuming that system relaxation is much faster than landscape deformation (adiabatic approximation), the extraction probability becomes 
\begin{equation}\label{eta_ramping1}
\tilde{\eta}(r)=\int_0^\infty \mathrm{d}F\frac{1}{r\tau_a(F)}\exp{-\int_0^F \mathrm{d}F'\frac{1}{r\tau(F')}},
\end{equation}
where $\tau(F)=[\tau_a^{-1}(F)+\tau_b^{-1}(F)]^{-1}$ is the mean lifetime of three-body complexes stretched by a force $F$ (see SI). Now that $\tilde{\eta}(r)$ depends on $\tau_a$ and $\tau_b$ separately, by pulling repeatedly at different loading rates, a cell may extract additional information about the free energy surface.   

In general, extraction under ramping force can be complex, depending on the entire distribution of rupture force, $p(F|r)$, at a given loading rate $r$:
\begin{equation}\label{eta_ramping}
\tilde{\eta}(r) = \int_0^{\infty}\mathrm{d}F \eta(F) p(F|r).
\end{equation}
However, if the rupture force distribution is relatively narrow compared to the variation of extraction probability with force, mean rupture force alone is predictive of extraction, i.e., $\tilde{\eta}(r)\approx \eta(\langle F \rangle_r)$, where $\langle F \rangle_r = \int_0^\infty \mathrm{d}F F p(F|r)$. Indeed, when plotting extraction against mean rupture force at varying loading rates ranging from $1$pN/s to $10^6$pN/s, data fall on theoretical curves under constant force (Fig.~\ref{fig2}A, squares falling on solid lines). Mathematically, this approximation holds as long as the difference between bond lengths, $|x_a^\ddagger-x_b^\ddagger|$, is small compared to the bond length per se (see SI).
The intuition is, $\eta$ varies slowly with force when bond lengths are similar, whereas $p(F|r)$ narrows quickly with increasing bond lengths.  

This simple relation, valid for realistic bond lengths of biomolecules, allows one to predict extraction under ramping force using constant-force theory. Conversely, it suggests a direct test of the constant-force theory using dynamic force spectroscopy, a powerful tool to extract kinetics of molecular transition in the absence of external forces from pulling adhesion bonds~\cite{dudko2008, neuman2008}. One way to establish the relation is to seek a match of success rate between constant-force and constant-speed pulling experiments, in which antigen fluorescence is tracked during repeated extraction attempts.

\begin{figure*}[thp]
\begin{center}
\includegraphics[width=0.9\textwidth]{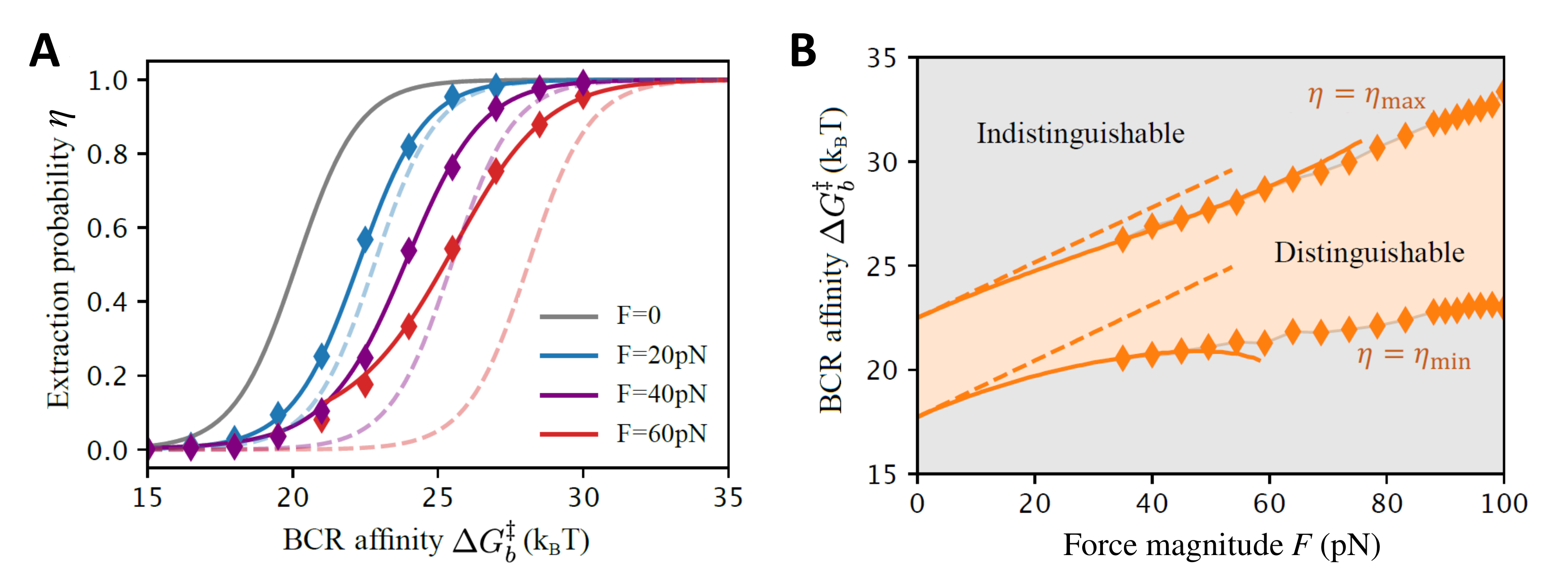}
\caption{
Tugging force stretches the extraction curve and expands the discrimination range.
(A) Extraction probability as a function of BCR affinity under different force magnitudes. Bell's model predicts a mere shift of the curve as force increases (dashed lines) since it neglects force-induced landscape deformation. Kramers theory properly accounts for such deformation and predicts stretching of the response curve, as verified by simulations (colored symbols, 200 runs each). 
(B) The range of distinguishable affinities (orange region), defined by $\eta(\Delta G_b^\ddagger; F)\in [\eta_\mathrm{min}, \eta_\mathrm{max}]$, expand as force increases, according to the landscape model (solid lines) and simulations (diamonds). Bell's model expects no expansion; dashed lines marking the range remain in parallel. 
$\eta_\mathrm{min}=0.1$, $\eta_\mathrm{max}=0.9$; $\Delta G_a^\ddagger=20\kbt, x_a^\ddagger=1.5$nm, $x_b^\ddagger=2$nm. 
}
\label{dis_curve}
\end{center}
\end{figure*}

Alternatively, one can reconstruct $\eta(F)$ under constant force based purely on rupture force histograms. 
With a single binding interface, bond lifetime under constant force can be expressed in terms of the rupture force distribution and loading rate~\cite{dudko2008} as 
$\tau_b(F) =\int_F^\infty p_b(f|r)\mathrm{d}f/(rp_b(F|r))$. Extending to three-body complexes, one needs to transform both rupture force distributions, $p_a(F|r)$ for the Ag-APC bond and $p(F|r)$ for the BCR-Ag-APC complex (Fig.~\ref{collapse} insets), to obtain the chance of extraction under constant force: 
\begin{equation}\label{reconstruct_eq}
\eta(F) = \frac{\tau(F)}{\tau_a(F)}=\frac{\int_{F}^\infty p(f|r)df}{\int_{F}^\infty p_a(f|r)df}\frac{p_a(F|r)}{p(F|r)}.
\end{equation}
This suggests that rupture force histograms measured at different loading rates should collapse onto a single master curve over a wide range of force magnitudes; this is indeed what we saw in Fig.~\ref{collapse}. Conversely, this relation allows one to extract intrinsic parameters of a multi-dimensional free energy profile, by fitting the force dependence of lifetimes from data collapse of rupture force histograms at multiple binding interfaces. Note that Eq.~\ref{reconstruct_eq} follows directly from the adiabatic approximation and is independent of the nature of the underlying free energy surface. In addition, this reconstruction works in the finite-barrier regime, which nicely encompasses the relevant range of rupture forces (10--40pN) observed in evolving B cells~\cite{natkanski2013}.

\subsection{Tugging force expands discrimination range}
We have demonstrated that pulling on a chain of molecular interactions causes differential effect on bond lifetimes and can potentially enhance discrimination stringency over force-free sensing schemes. But, in order to rank B cells with varying affinities, discrimination must extend over a wide dynamic range. To what extent can tugging forces influence the operation range of affinity-dependent extraction?

Our theory, supported by Brownian dynamics simulations, shows that the extraction curve interpolates smoothly from none to all acquisition as BCR affinity increases (Fig.~\ref{dis_curve}A). The discrimination range can, therefore, be defined as the affinity span between almost vanishing ($\eta_\mathrm{min}$) and nearly full ($\eta_\mathrm{max}$) extraction, the limits where sensitivity to affinity changes is lost. Fig.~\ref{dis_curve}A presents extraction curves under different force magnitudes and Fig.~\ref{dis_curve}B extracts the discrimination range (orange region) over wide-ranging pulling strengths. 

We find that as force increases, not only that the extraction curve shifts toward higher affinities, like it does in Bell's model (Fig.~\ref{dis_curve}A, dashed lines), but it also stretches such that relatively low affinities remain distinguishable (Fig.~\ref{dis_curve}A, solid lines), thereby substantially broadening the dynamic range (Fig.~\ref{dis_curve}B, delineated by solid lines and symbols). We argue that this response-curve stretching reflects affinity-dependent force effect resulting from landscape deformation. 
From Eq.~\ref{barrier} we see that force-induced barrier reduction has a nonlinear offset that depends inversely on the intrinsic BCR-Ag affinity, $\Delta G_b^\ddagger$. In other words, if the intrinsic barrier to BCR-Ag bond rupture were lower, pulling force would cause a smaller barrier reduction. As a result, as pulling applies, B cells of lower affinity can maintain the extraction level via a smaller increase in affinity: 
$\Delta G_b^{\ddagger}(\eta; F)-\Delta G_b^{\ddagger}(\eta; 0)\approx F(x_b^\ddagger-x_a^\ddagger)-(1/2)(1-v)\left[(Fx_b^\ddagger)^2/\Delta G_b^\ddagger(\eta;0)-(Fx_a^\ddagger)^2/\Delta G_a^\ddagger\right]$.
To directly demonstrate the range expansion, we find the distinguishable range, $\Delta\Delta G_b^\ddagger\equiv\Delta G_b^\ddagger(\eta_\mathrm{max})-\Delta G_b^\ddagger(\eta_\mathrm{min})$, to the leading order in force magnitude  
\begin{equation}
\frac{\Delta\Delta G_b^\ddagger(F)}{\Delta\Delta G_b^\ddagger(0)} \approx 1+\frac{1-v}{2}\frac{\left(Fx_b^\ddagger\right)^2}{\Delta G_b^\ddagger(\eta_\mathrm{max};0)\Delta G_b^\ddagger(\eta_\mathrm{min};0)}.\nonumber
\end{equation}
Note that only force-induced stretching that begins at the quadratic order contributes to range expansion, while the linear term that only shifts the curve drops off, explaining why Bell's model that neglects landscape deformation expects no range expansion (Fig.~\ref{dis_curve}B dashed lines).

Intriguingly, under dynamic ramping forces, the expansion of discrimination range is even more pronounced, because mean rupture force increases with BCR affinity (Fig.~S4). This can be understood from $\tilde{\eta}(r;\Delta G_b^\ddagger)=\int_0^\infty\mathrm{d}F p(F|r,\Delta G_b^\ddagger)\eta(\Delta G_b^\ddagger,F)\approx\eta(\langle F\rangle_{r,\Delta G_b^\ddagger})$. It says that, for a given loading rate $r$, as $\Delta G_b^\ddagger$ increases, a stronger mean rupture force $\langle F\rangle_{r,\Delta G_b^\ddagger}$ results in a stronger suppression on extraction, which further flattens the response curve, raises the affinity ceiling, and widens the discrimination range.  
This implies an interesting possibility: low-affinity B cells apply small forces to extract antigen whereas higher-affinity B cells use stronger forces. This affinity-dependent force application is predicted to further broaden distinguishable affinities.

How much range expansion can be expected based on realistic parameters?
As illustrated in the example of Fig.~\ref{dis_curve}, discrimination range expands from $5\kbt$ (150-fold increase in the binding constant) to $10\kbt$ (20000-fold increase), as force increases from 0 to 100pN. 
Affinity maturation is known to achieve up to $10^3-10^4$ fold affinity increase \textit{in vivo}~\cite{eisen1964}; our model suggests that this is plausible when B cells use moderate pulling forces.

\subsection{Antigen extraction via molecular tug of war raises affinity ceiling and accelerates adaptation}
We have shown that physical extraction of antigen relates conformational changes to rupture kinetics and allows a gradual dependence of signal acquisition on BCR affinity over a wide dynamic range. But whether, and how, does active sensing by individual cells influence adaptation of a polyclonal population? An ultimate test of plausible physical behaviors is to subject the resulting phenotype to natural selection. Combining intravital imaging and single-cell sequencing, researchers have been able to track the ancestry of proliferating cells on the move~\cite{tas2016}. They found that the reproductive fitness of a B cell is proportional to the amount of antigen it acquires from the APC and subsequently presents to the helper T cell~\cite{gitlin2014}, suggesting a link from receptor affinity to clonal fitness via antigen extraction efficiency. We thus propose that B cells may do mechanical work to drive their own evolution. To test this hypothesis, we couple the physical theory of antigen extraction to a minimal model of affinity maturation, simulate ensembles of cell populations pulling at different strengths, and examine features of adaptive dynamics. 

\begin{figure*}[htp]
\begin{center}
\includegraphics[width=0.93\textwidth]{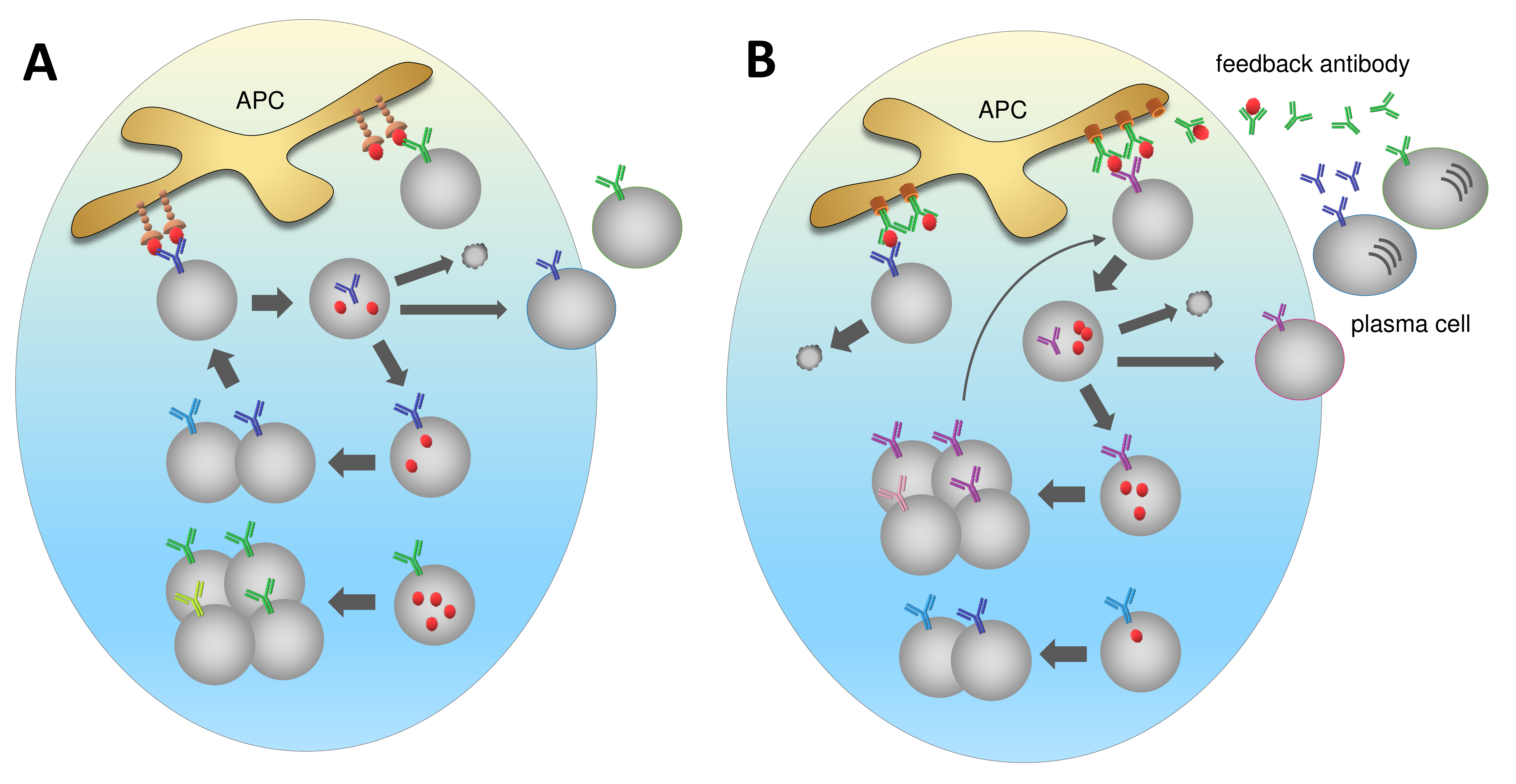}
\caption{A schematic of GC reaction with fixed versus renewable antigen tethers illustrates the impact of antibody feedback. Cycles of antigen extraction, death, differentiation/recycle, replication and mutation alter the composition and size of a B cell population over time. (A) If tethers are fixed, higher-affinity clones (green) will likely extract a larger amount of antigen (red dot) and produce more offspring, expanding in size at the expense of lower-affinity clones (blue). Hence, population affinity increases but eventually hits a ceiling, once all clones carry BCRs stickier than the fixed tether and efficiently acquire antigen. (B) Tethers are constantly updated by antibodies secreted by most potent plasma cells available, as they compete better for antigen binding and presentation on the APC. This causes a steadily elevated selection pressure: Clones with inferior or similar affinity to the tether are likely to lose the tug of war and die (blue). More potent clones (magenta), instead, will likely win over the tether (secreted by green cells), acquire antigen, and differentiate into plasma cells that supply feedback antibodies as tethers in subsequent GC cycles. As a result, affinity ceiling is lifted; sustained adaptation results from antibody feedback.}
\label{GCR}
\end{center}
\end{figure*}

Specifically, we implemented a birth-death-mutation model of GC reaction using agent-based simulations (Fig.~\ref{GCR}; SI text and Table I), which implements cycles of antigen extraction, death, differentiation/recycle, proliferation and mutation that drives stochastic clonal expansion and an overall increase in affinity. The key ingredient is an affinity-dependent proliferation rate, where our tug-of-war model bridges BCR affinity and Ag extraction efficiency which, in turn, determines clonal fitness. To be concrete, we assume a sigmoidal dependence of clonal fitness $\lambda_i$ on extraction probability $\eta_i$ on top of a logistic growth (time $t$ being in discrete GC reaction cycles): 
\begin{equation}\label{lambda_t}
\lambda_i(t)=\lambda_0\frac{\eta_i(t)}{\eta_0+\eta_i(t)}\left(1-\frac{N(t)}{N_c}\right),
\end{equation} 
where $i$ indexes cells and $N_c$ denotes an overall carrying capacity that accounts for space and resource limitation. The parameter $\eta_0$ represents the extraction level at half-maximum growth rate and sets an effective threshold of clonal survival. We obtain $\eta$ from the analytical theory and keep track of simulated population dynamics and affinity trajectories of surviving cells.   

\begin{figure*}[htp]
\begin{center}
\includegraphics[width=1\textwidth]{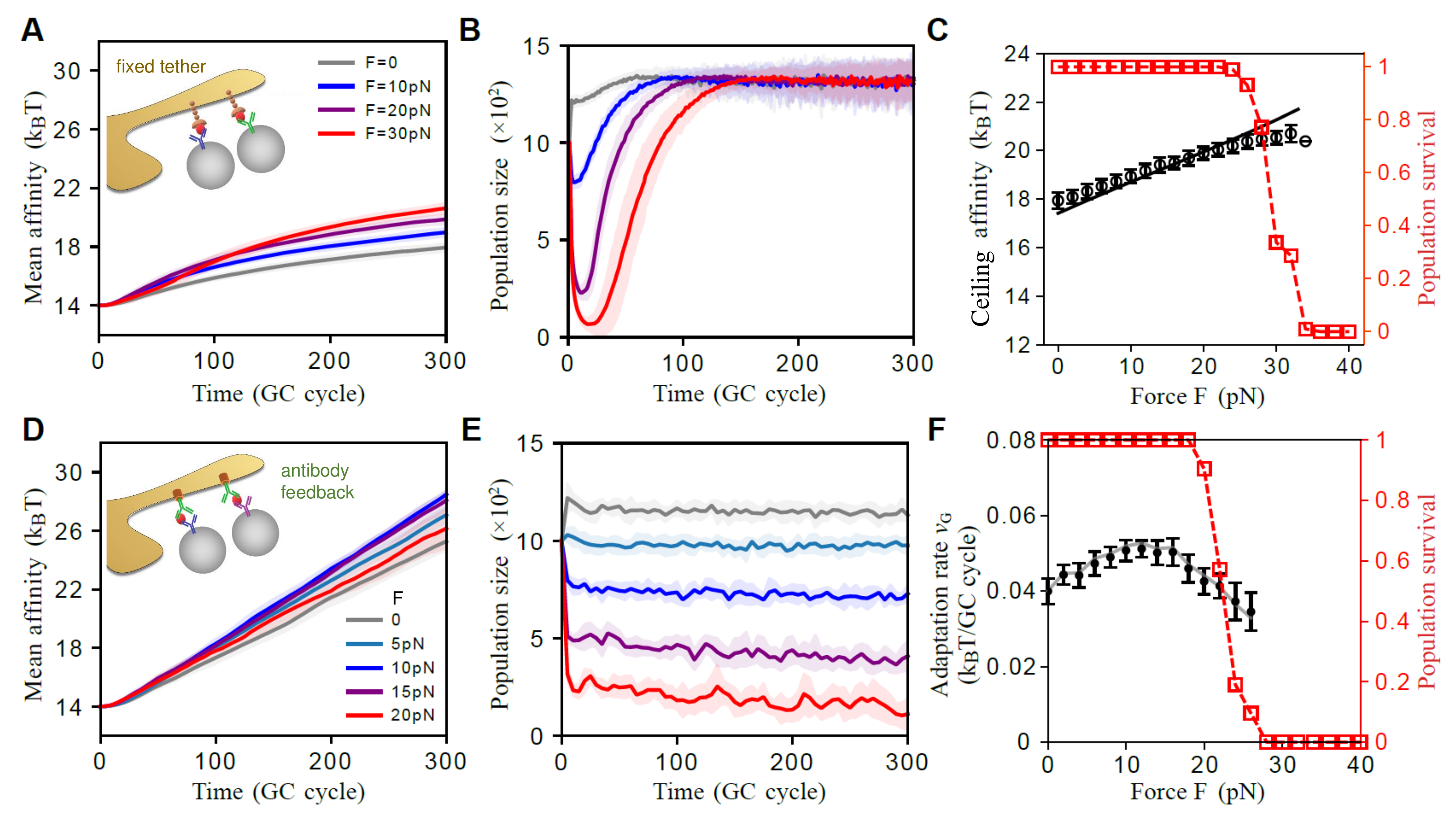}
\caption{
Optimal tugging forces balance the quality and magnitude of emergent responses.
Time trajectories of population-mean affinity and population size, along with force dependence of the ceiling affinity and adaptation rate are presented for fixed (upper row) and renewable (lower row) antigen tethers.   
(A, B) For fixed tethers, mean affinity approaches saturation (A) as population size recovers to a force-independent steady level following an early bottleneck (B). 
(C) Stronger forces raise the affinity ceiling (black symbols: mean affinity at the end of 300 GC cycles from 100 simulations; solid line: analytical prediction based on $\eta(\Delta G_b^\ddagger, F)=0.96$). Too strong pulling leads to a rapid fall in population survival (red symbols: fraction of surviving populations; dashed line is to guide the eye). 
(D, E) With feedback antibodies renewing the tethers, mean affinity exhibits a steady increase (D) and population size stabilizes to force-dependent values (E).
(F) The rate of affinity increase, $v_G$, shows a non-monotonic dependence on force (black symbols: mean adaptation rate over the last 200 GC cycles from 100 runs; solid line: prediction from Eq.~\ref{price}). Population survival shows a similar rapid decline as in panel (C), but starting at a lower pulling strength. In panels A, B, D and E, each solid line represents an average over 100 runs with the shade indicating variation among runs. Simulation steps and parameters are provided in SI. 
}
\label{evo_fig}
\end{center}
\end{figure*}

A variety of tethering receptors can present antigen on the surface of APCs.
If the antigen tether remains unchanged (e.g. Fc$\gamma$ and CR2 receptors; Fig.~\ref{GCR}A), as we have so far assumed, population mean affinity first increases and then levels off (Fig.~\ref{evo_fig}A). Concomitantly, population size first falls and, if successfully rescued by clones acquiring beneficial mutations, subsequently recovers to a force-independent steady size as $\eta$ approaches saturation (Fig.~\ref{evo_fig}B). In this case, pulling-induced stretching of the extraction curve promotes discrimination of strong affinities while facilitating survival of lower affinity clones. These effects combine to sustain adaptation and support clonal diversity at once. As a result, as force increases, the affinity ceiling rises; ceiling affinity matches the prediction based on vanishing discrimination (Fig.~\ref{evo_fig}C, black solid line). Yet, stronger pulling also increases the risk of population extinction due to a deeper bottleneck (Figs.~\ref{evo_fig}B and ~\ref{evo_fig}C). 
Note that accounting for landscape deformation under pulling (nonlinearity captured by Kramers theory) is key to correctly predicting ceiling affinity (Fig.~S6), manifesting macroscopic impact of microscopic characteristics.

Interestingly, the tug-of-war configuration naturally supports inter-generational feedback via antibodies (Fig.~\ref{GCR}B). In fact, antibodies with improved affinities, secreted by newly differentiated plasma cells, may preferentially present antigens on the APC in the form of immune complexes~\cite{zhang2013, zhang2016, toellner2018}. In this case, population mean affinity increases at a steady rate (Fig.~\ref{evo_fig}D); population size stabilizes to a force-dependent steady level (Fig.~\ref{evo_fig}E), since $\eta$ decreases with increasing pulling strength (Fig.~S7B). Notably, the adaptation rate, $v_G$, i.e., the steady rate of increase in mean affinity, exhibits a non-monotonic dependence on force magnitude, peaked around $F=$10--20pN (Fig.~\ref{evo_fig}F, black symbols). 

This behavior can be understood qualitatively through 
the conceptual picture of beneficial mutation-selection balance in asexual populations~\cite{desai2007}.
With antibody feedback, the same affinity-increasing mutations not only yield new variants that advance the affinity distribution (pulling at the leading edge), but they also produce antibodies that compete favorably in presenting antigen and hence impede selection of lower-affinity clones (pushing at the rear). Consequently, the affinity distribution proceeds at a steady speed. Once feedback antibodies establish a steady affinity gap between the BCR and antigen tether, $\eta$ becomes stable, followed by population size and selection strength, hence a steady $v_G$.   
As force increases, $\eta$ decreases hence population size falls while selection strength rises. This competition results in maximum adaptation rates at intermediate pulling forces. 

A more quantitative expectation, especially for the force dependence of adaptive dynamics, can be made based on the Price equation~\cite{price1970}, which relates the adaptation rate $v_G$ to the covariance of fitness and the trait of interest. In the context of affinity maturation, it states that (see SI text)
\begin{equation}\label{price}
v_G\equiv 
\langle\dv{\overline{\Delta G_b^\ddagger}}{t}\rangle=\langle \text{Cov}(\Delta G_b^\ddagger, \lambda)/\overline{\lambda} \rangle
\approx \langle \alpha(\overline{\Delta G_b^\ddagger}){\rm Var}(\Delta G_b^\ddagger) \rangle,
\end{equation}
where overbars denote population mean and angular brackets stand for ensemble average; covariance and variance are taken with respect to a single population. 
Here, the discrimination stringency $\alpha\equiv\mathrm{d}\ln{\lambda}/\mathrm{d}\Delta G_b^\ddagger$ characterizes how sensitively fitness responds to affinity changes and thus quantifies selection strength.
The approximate relation holds when fitness varies gently over the width of the affinity distribution. At non-extreme magnitudes, stronger puling first accelerates adaptation by enhancing selection, but then slows it as affinity variance falls with shrinking population size, hence the non-monotonic dependence of $v_G$ on force. 
Different from antigen masking that reduces the stimuli of GC reaction and ends AM~\cite{zhang2013}, our work suggests an alternative role of antibodies as renewable tethers: increasingly stickier antibodies from recent generations hold antigens on the APC, maintaining selection pressure without causing population extinction. This proposal is supported by \textit{in vivo} studies in which passively injected antibodies of higher affinities replace endogenous antibodies to present antigens~\cite{zhang2016} and enhance response quality~\cite{garg2019}. Moreover, the predicted $v_G$ ($\sim0.05k_\mathrm{B}T$ per GC cycle) matches the observed rate of affinity increase~\cite{tas2016} (SI Table II and references therein).

Further, Fig.~\ref{evo_fig} suggests that this feedback mechanism operates to keep the cell population between two absorbing states, namely, extinction (under too much pressure) and saturation (due to ineffective selection). This desired regime of persistent adaptation can be anticipated theoretically: From $\mathrm{d}\eta/\mathrm{d}t=\left(\partial \eta/\partial\Delta G_b^\ddagger\right)\mathrm{d}{\Delta G_b^\ddagger}/\mathrm{d}t+\left(\partial \eta/\partial\Delta G_a^\ddagger\right)\mathrm{d}{\Delta G_a^\ddagger}/\mathrm{d}t+\left(\partial \eta/\partial F\right)\mathrm{d}{F}/\mathrm{d}t$, where $\partial\eta/\partial\Delta G_b^\ddagger>0$, $\partial\eta/\partial\Delta G_a^\ddagger<0$, and $\partial\eta/\partial F<0$, we see that as intrinsic affinity $\Delta G_b^\ddagger$ increases through affinity maturation, extraction elevates toward saturation. Two processes can counterbalance it by ramping up negative feedback: (i) enhancing tether affinity and (ii) upregulating pulling force. When ramping at a rate that matches the pace of AM, these processes effectively create a moving frame in which affinity distribution stands still and extraction remains away from saturation. Optimal ramping rates thus result.
Intriguingly, cells can implement (i) through antibody feedback and realize (ii) via signaling that instructs assembly of additional actomyosin bundles as more BCR-Ag-APC bound complexes form. Once $\eta$ becomes steady, population size and adaptation rate would follow. 
The timing and rate of steady adaptation can be modulated by the efficiency and strength of feedback (Fig.~S8).

Therefore, both in the presence and absence of antibody feedback, a favorable level of mechanical energy expense is bounded from above by a minimum amount of extracted antigen required for cell activation and population survival (Figs.~\ref{evo_fig}C and ~\ref{evo_fig}F, red symbols). In this sense, the optimal magnitude of contractile forces acting on single receptors might be selected on the population level, balancing the needs to adapt rapidly and sustainably. Indeed, the predicted force magnitudes for efficient adaptation (10--20pN) match the range of rupture forces measured by single-molecule pulling experiment on tethered antibodies (10--40pN)~\cite{natkanski2013} and via tension sensors in live B cells (above 9pN)~\cite{nowosad2016}. While demonstrated using the adaptive immune system as an example, our results point to a broader picture in which active sensing of cells physically modulates selection pressures on molecular recognition.

\section{Discussion}

We have shown, in line with experiments on multiple scales, that active pulling forces regulate physical extraction of antigen via a molecular tug of war, which couples internal dynamics of cells to the mechanical environment, thus allowing comparative measurement (via competitive bond rupture) and dynamic feedback (via renewable antigen tethers). Employing slip bonds that dissociate faster under force, evolving B cells pull vigorously against stiff APCs to enhance the contrast between similar affinities at a cost of the absolute level of extraction. Force-induced stretching of response curves expands the range of discrimination. However, too strong pulling may risk extinction of an evolving cell population, thereby limiting evolvable affinities. Interestingly, the magnitude of force applied on the molecular scale appears to be selected on the organismal level, for a balance between the quality and magnitude of collective responses. 

Predictions of our theory can be tested using dynamic force spectroscopy combined with live-cell imaging. The extraction curve, $\tilde{\eta}(r)\approx \eta(\langle F \rangle_r)$, if able to collapse the dynamic-force data onto the constant-force theory (Fig.~\ref{fig2}A), will aid in understanding rupture dynamics and predicting extraction propensity based on mean rupture force; a smaller difference in stiffness between the tugging and tethering complexes is expected to improve the match. An independent test of the constant-force theory is to reconstruct $\eta(F)$ from rupture force histograms obtained over a wide range of loading rates in single-molecule pulling experiment (Eq.~\ref{reconstruct_eq}, Fig.~\ref{collapse}). In both cases, $\eta$ can be estimated by counting successful events out of many extraction attempts; success is determined by tracking antigen fluorescence during rupture.
Importantly, by fitting data to the analytical theory, one can extract intrinsic parameters characterizing the multi-dimensional binding landscape, especially the force-free strength of antigen tether that would otherwise be hard to measure.   
Lastly, data collapse for $\eta(F)$ can be used to probe landscape complexity: deviation from theory might indicate the presence of more than one attractors. 

Pulling against adhesive bonds is reminiscent of mechanosensing by diverse cell types. A classic example is the catch bond behavior, i.e. prolonged bond lifetimes under force, of immune T cells as a means of proofreading for self-foreign discrimination and that of leukocytes for rolling on vessel walls near a wound or an infection site~\cite{marshall2003}. Different from this usual form of adhesive selection, B cells employ slip bonds that dissociate faster under force. We show that the slip-bond characteristic underlies the distinct behaviors of naive and evolving B cells: evolving cells exert strong forces against stiff APCs and acquire fewer antigens than naive cells that use weak forces, trading absolute activation signal for enhanced affinity discrimination. 

In this work, we have focused on how force modulates dissociation of individual receptor-antigen-tether complexes, treating extraction events as independent. Experiments showed that antigen extraction occurs through pulling on clusters at the B cell-APC interface, following the formation of a multifocal contact pattern~\cite{nowosad2016}. Accounting for coupling of rupture events by membrane deformation and/or load sharing would be an interesting future direction, as it will reveal cooperative mechanisms that may enhance discrimination accuracy and broaden the dynamic range on the cellular level. In addition, this study can shed light on why evolving cells limit the size of BCR-Ag clusters and whether this behavior places additional constraints on the favorable force range.

Although our treatment of antigen extraction is mean-field in nature, it reveals that, even with independent receptors, immune cells can already tune selection pressure through multiple pathways, such as renewing antigen tethers with newly produced antibodies and augmenting the pulling machinery via enhanced signaling, as receptor quality steadily improves. These feedback mechanisms, alone or combined, act to suppress extraction and cell activation and hence sustain adaptation. Moreover, physical environments can adaptively change to impact selection. For instance, the ability of APCs to modulate their stiffness in response to inflammatory signals~\cite{bufi2015}, in combination with the stiffness sensitivity of immune cells, may support proper responses to complex signals at different developmental stages.  
Therefore, understanding these adaptive pathways will uncover novel strategies to modulate B cell selection by manipulating the internal drive (e.g. altering myosin motor activity with drugs), modifying the physical environment, and controlling the manner of antigen presentation. 

This work joins the large efforts in providing a physical basis for biological recognition~\cite{george2017, dudko2016, altan2020, endres2008, mehta2009, mora2010, skoge2011, haselwandter2014, persat2015, beroz2017, rozycki2010, ten2016, belardi2020}. While sharing the spirit of linking conformational changes to kinetics~\cite{savir2007}, here the link is provided not by binding equilibrium but by off-equilibrium rupture dynamics at multiple binding interfaces --- an activity-dependent process that causes differential deformation of the underlying free energy landscapes. Often, higher energy dissipation is associated with enhanced accuracy of sensory functions and parameter estimates performed by biological systems. Our work shows that, counterintuitively, discrimination capacity is maximal at moderate energy expenses, if energy dissipation results in a negative feedback on the activation signal. Therefore, more broadly, our analysis highlights the need for understanding the connection between the observed phenotype and the underlying physical dynamics, in order to infer key constraints and determinants of system adaptability.

\section{Acknowledgement}
We thank Tom Chou for valuable discussions and funding support via NIH R01HL146552 at the consolidation stage of this project. This work is in part supported by the Bhaumik Institute for Theoretical Physics at UCLA and an NSF CAREER Award PHY-2146581 to SW.
\clearpage

\newpage
\bibliographystyle{unsrt}
\bibliography{main}

\end{document}